\documentclass[a4paper,12pt]{article}
\usepackage{graphicx}
\usepackage[T1,T2A]{fontenc}
\usepackage[utf8]{inputenc}
\usepackage[english]{babel}
\usepackage{amssymb}
\usepackage{amsmath}
\usepackage{setspace}
\usepackage{xcolor}
\usepackage{wrapfig}
\graphicspath{{figures/}}
\begin{document} 

	\begin{center}
		{\large
			{\bf 
				Thermodynamic and magnetic properties of the Ising model with nonmagnetic impurities
			} 
		}
		\vskip0.5\baselineskip{
			\bf 
			
			D.~N.~Yasinskaya$^{1}$,
			Yu.~D.~Panov$^{1}$
		}
		\vskip0.5\baselineskip{
			$^{1}$Ural Federal University, 19 Mira street, 620002 Ekaterinburg, Russia
		}
	\end{center}
	
	\begin{center}
		\textbf{Annotation}
	\end{center}
	
	We consider a system of Ising spins $s=1/2$ with nonmagnetic impurities with charge associated with pseudospin $S=1$. The charge density is fixed pursuant to the concentration $n$. Analysis of the thermodynamic properties in the one-dimensional case showed the presence of so-called pseudotransitions at the boundaries between the staggered charge ordering and (anti)ferromagnetic ordering. In the case of n=0, a ``1st order'' pseudotransition was discovered. This type of pseudotransition is inherent for a series of other one-dimensional frustrated models. However, for $n \neq 0$ we discovered a new type of ``2nd order'' pseudotransition, which had not previously been observed in other systems.\\
	
	This work was supported in by the Ministry of Education and Science of the Russian Federation (project FEUZ-2023-0017)
	
	\section{Introduction}
	
	Low-dimensional magnetic structures can demonstrate long magnetization relaxation times due to a combination of uniaxial anisotropy and strong magnetic interactions. Thereby such magnets at finite temperatures find use in applications of quantum computing [1] and spintronics.
	
	The presence of anisotropy and frustration leads to the rich phase diagrams and to such unusual phenomena as magnetic plateaus [2], quasi-phases and pseudotransitions [3], as well as an increase in the magnetocaloric effect [4]. Disorder also significantly affects the phase, critical and magnetic properties of systems, and is also a source of frustration. The fulfillment of the Rojas criterion [5] for frustrated phases can lead to the presence of such a subtle pseudocritical phenomenon as pseudotransitions. They are accompanied by an abrupt change in the type of disordered state of the system, in which some thermodynamic functions exhibit very sharp features, although they remain continuous.
	
	\section{Model and methods}
	
	We use the transfer-matrix method to investigate the 1D Ising model with interacting nonmagnetic impurities. The Hamiltonian of the system has the following form:
	
	\begin{equation}
		\label{H}
			H = \Delta \sum_i S_{i,z}^2 + V \sum_{\langle ij \rangle} S_{i,z}S_{j,z} + J  \sum_{\langle ij \rangle} \sigma_{i,z} \sigma_{j,z} - h  \sum_i \sigma_{i,z} - \mu  \sum_i S_{i,z},
	\end{equation}
	where $S=1$ is the Ising pseudospin associated with the charge degrees of freedom; $\sigma=1/2$ is the Ising magnetic spin; $\Delta$ is the on-site density-density correlations in the form of single-ion anisotropy; $V$ is the inter-site density-density correlations in the form of the Ising exchange interaction; $J$ is the spin exchange coupling; $h$ is the external magnetic field; $\mu$ is the chemical potential, which is used to fix the total charge density in the system: $nN=\sum\limits_{i=1}^N S_{i,z}$.
	
	The transfer matrix for the Hamiltonian~(\ref{H}) is
	
	\begin{equation}
		\hat{T} = \left(
		\begin{array}{cccc}
			g^2 x^2 y & \frac{x^2}{y} & g k x & \frac{g x}{k} \\
			\frac{x^2}{y} & \frac{x^2 y}{g^2} & \frac{k x}{g} & \frac{x}{g k} \\
			g k x & \frac{k x}{g} & k^2 z & \frac{1}{z} \\
			\frac{g x}{k} & \frac{x}{g k} & \frac{1}{z} & \frac{z}{k^2} \\
		\end{array}
		\right),
	\end{equation}
	where $x = e^{-\frac{\beta \Delta }{2}}$, $y = e^{-\beta V}$, $z = e^{-\beta J}$, $k = e^{\beta \frac{h}{2}}$, $g = e^{\beta \frac{\mu }{2}}$, and $\beta = \frac{1}{T}$.
	
	According to the Perron-Frobenius theorem, the transfer matrix will have only one largest eigenvalue $\lambda_1$. In this case, the partition function built on the eigenvectors of the transfer matrix will have the following form in the thermodynamic limit:	
	\begin{equation}
		Z = Sp(\hat{T}^N) = \lambda_1^N + \lambda_2^N + \lambda_3^N + \lambda_4^N \underset{N \rightarrow \infty}{\longrightarrow} \lambda_1^N.
	\end{equation}
	It means that the grand potential, as well as other thermodynamic quantities will depend only on the $\lambda_1$.
	
	\section{Results}
	
	Analysis of the ground state of the system showed the presence of eight non-trivial phases. Ferromagnetic (FM, FM+PS) and antiferromagnetic (AFM, AFM+PS) phases comprise macroscopic phase separation with charge droplet and zero residual entropy. The remaining six phases have non-zero residual entropy, depending on the charge density $n$, which means they are frustrated phases.
	
\begin{figure}[h!]
		\centering
		\begin{minipage}{0.49\linewidth}
			\centering
			(a) $\vert h \vert \leq 2V$
			\includegraphics[width=\linewidth]{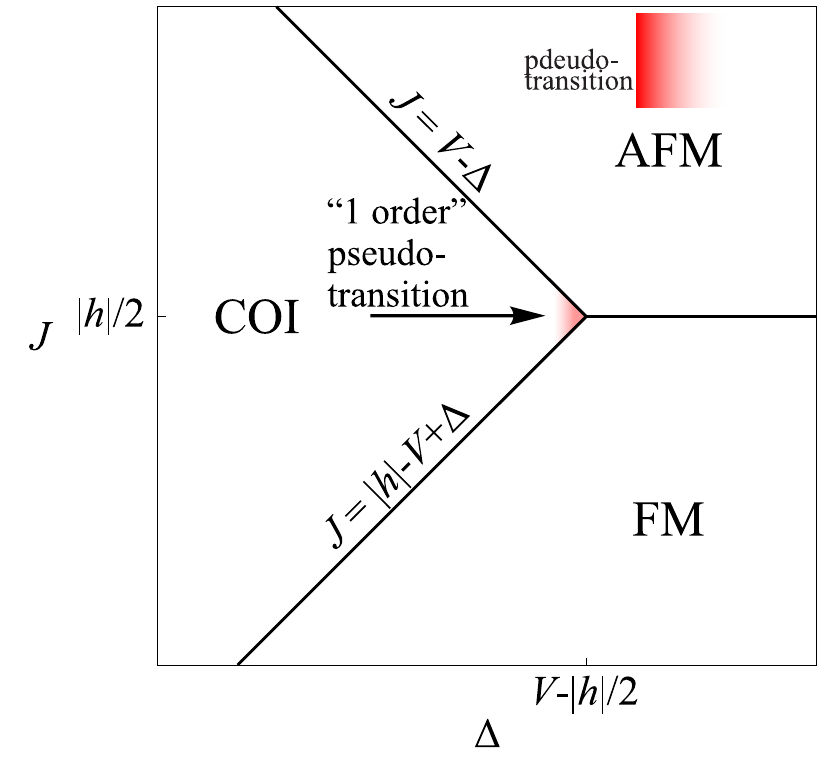}
		\end{minipage}
		\hfill
		\begin{minipage}{0.49\linewidth}
			\centering
			(b)  $\vert h \vert >2V$
			\includegraphics[width=\linewidth]{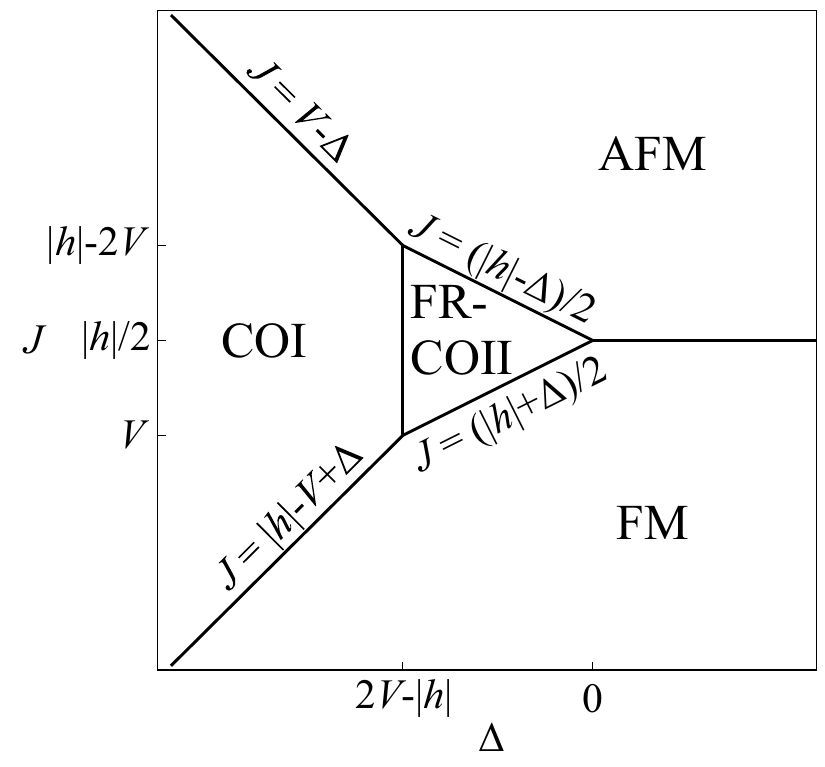}
		\end{minipage}
	\includegraphics[width=\linewidth]{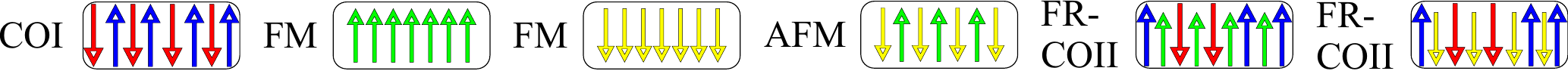}
	\caption{Ground state phase diagrams in the ($J$,$\Delta$) plane for a pure spin chain, $n=0$, in (a) weak magnetic field $\vert h \vert \leq 2V$; (b) strong magnetic field $\vert h \vert >2V$}
	\label{fig1}
	\end{figure}

	The ground state phase diagrams for the pure system ($n=0$) are presented in Figure~\ref{fig1}. In addition to the FM and AFM phases, a checkerboard charge order COI and a phase with alternating charges of two types and spins FR-COII are formed. Near the boundary between COI and the frustrated mixture of FM and AFM phases, the Rojas criterion [5] is satisfied, which indicates the presence of pseudotransitions in the system.
	
\begin{figure}[h!]
	\centering
	\begin{minipage}{0.49\linewidth}
		\centering
		(a) Entropy
		\includegraphics[width=\linewidth]{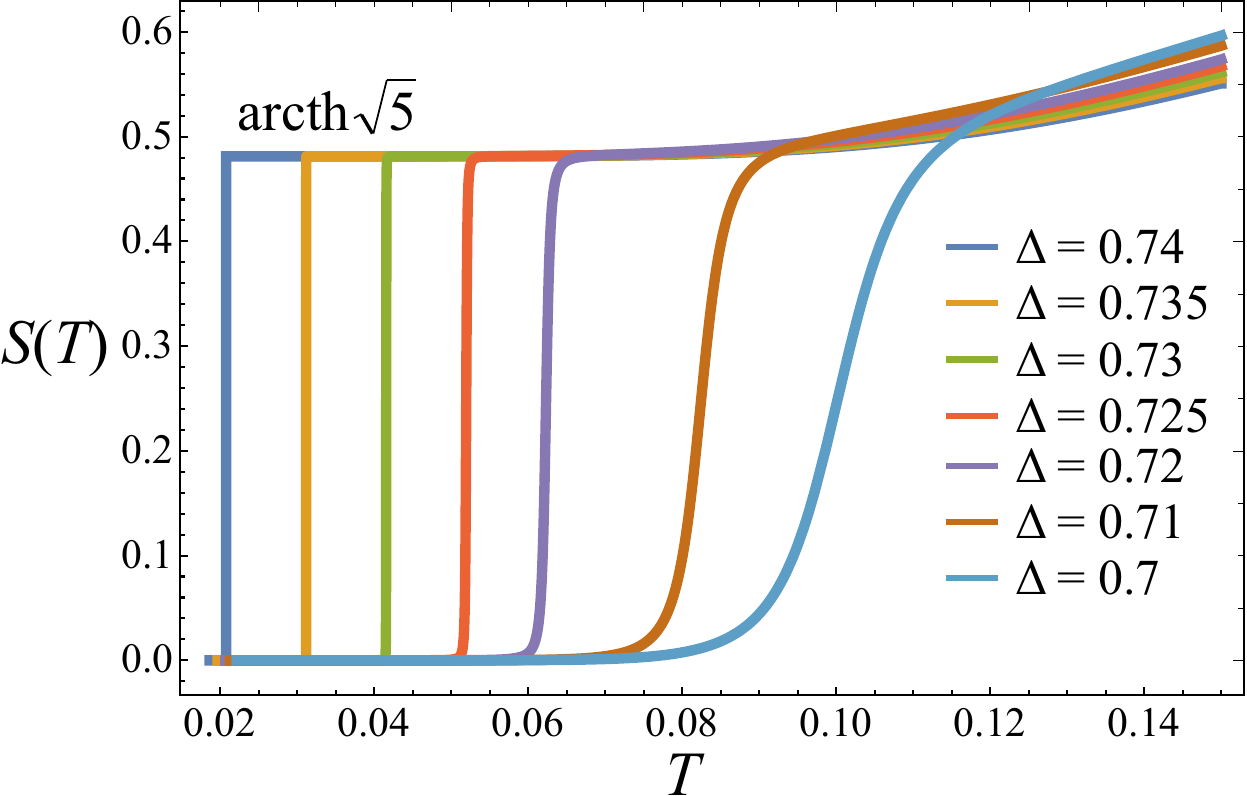}
	\end{minipage}
	\hfill
	\begin{minipage}{0.49\linewidth}
		\centering
		(b) Specific heat
		\includegraphics[width=\linewidth]{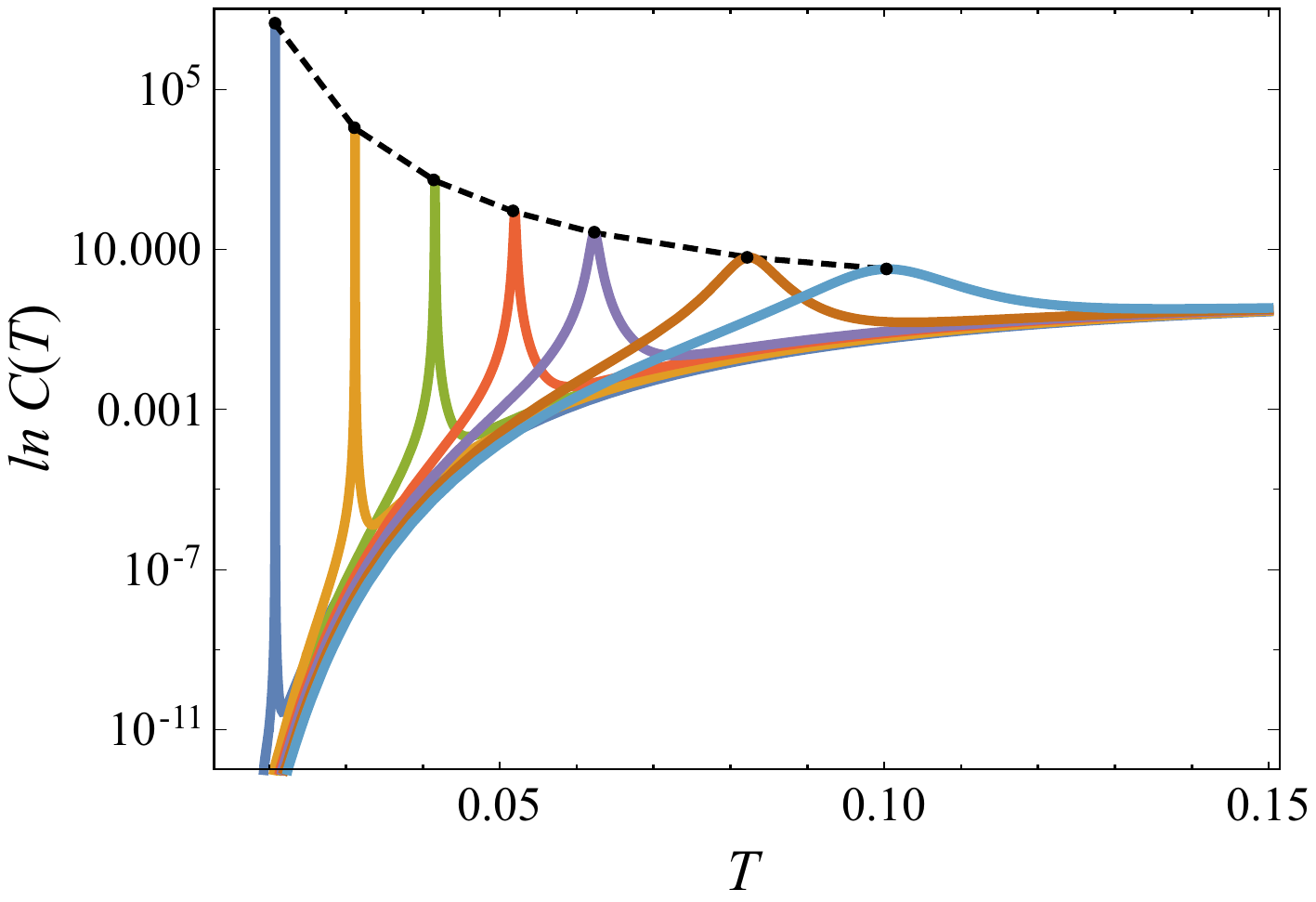}
	\end{minipage}
	\caption{Temperature dependences of (a) entropy and (b) specific heat demonstrate pseudotransitions from a frustrated mixture of FM and AFM quasi-phases to the COI quasi-phase with zero entropy}
	\label{fig2}
\end{figure}

Figure~\ref{fig2} shows the temperature dependences of entropy and specific heat near the boundary between COI and AFM+FM, where pseudotransitions can be observed. Entropy (as well as other first derivatives of the grand potential like magnetization) has a sharp jump (similar to a discontinuity). Specific heat (as well as other second derivatives of the grand potential like magnetic susceptibility and correlation length) has a giant peak (similar to singularity). Nevertheless, all thermodynamic quantities remain analytical functions. Evaluation of the pseudocritical exponents yielded $\alpha=3$, $\gamma=3$ and $\nu=1$, which correspond to the exponents in other frustrated models with pseudotransitions.

\begin{figure}[h!]
	\centering
	\begin{minipage}{0.49\linewidth}
		\centering
		(a) $0 < n < 1/2$
		\includegraphics[width=\linewidth]{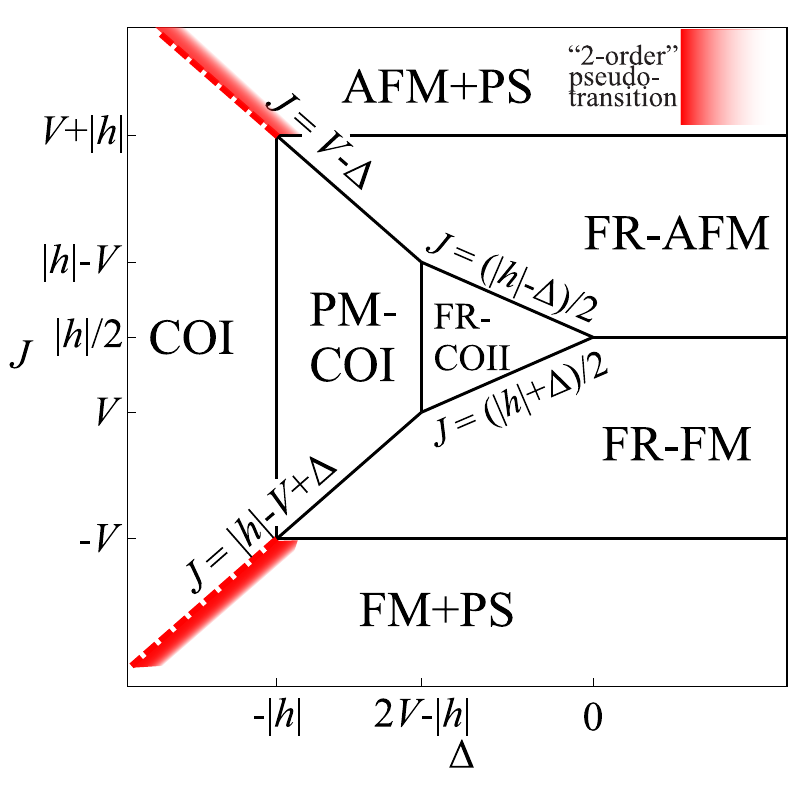}
	\end{minipage}
	\hfill
	\begin{minipage}{0.49\linewidth}
		\centering
		(b) $n \geq 1/2$
		\includegraphics[width=\linewidth]{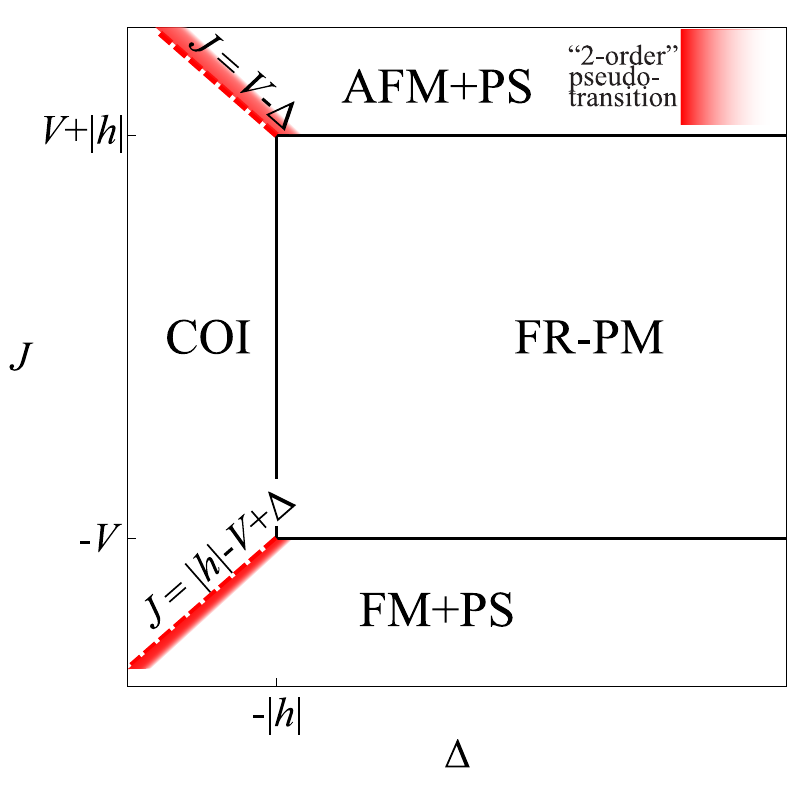}
	\end{minipage}
	\includegraphics[width=\linewidth]{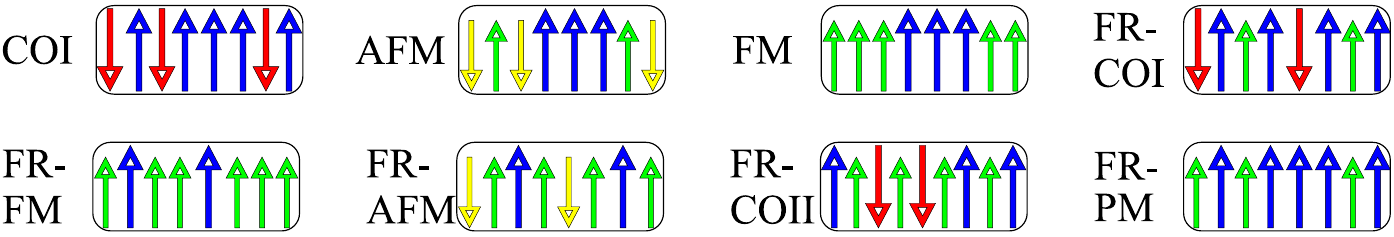}
	\caption{Ground state phase diagrams in the ($J$,$\Delta$) plane for $|h|>2V$ for (a) weakly diluted spin chain, $0<n<1/2$; (b) strongly diluted spin chain, $n \geq 1/2$.}
	\label{fig3}
\end{figure}

Figure~\ref{fig3} shows ground state phase diagrams in the case when charge density $n$ is non-zero. Now the magnetic phases comprise phase separation (PS), and many frustrated phases are added. They include the magnetic phases with randomly distributed impurities (FR-FM, FR-AFM), the frustrated paramagnetic phase (FR-PM), and also staggered charge phase with paramagnetic response (PM-COI). Regions where pseudotransitions are observed are also highlighted in red.

\begin{figure}
	\centering
	\begin{minipage}{0.49\linewidth}
		\centering
		(a) Entropy
		\includegraphics[width=\linewidth]{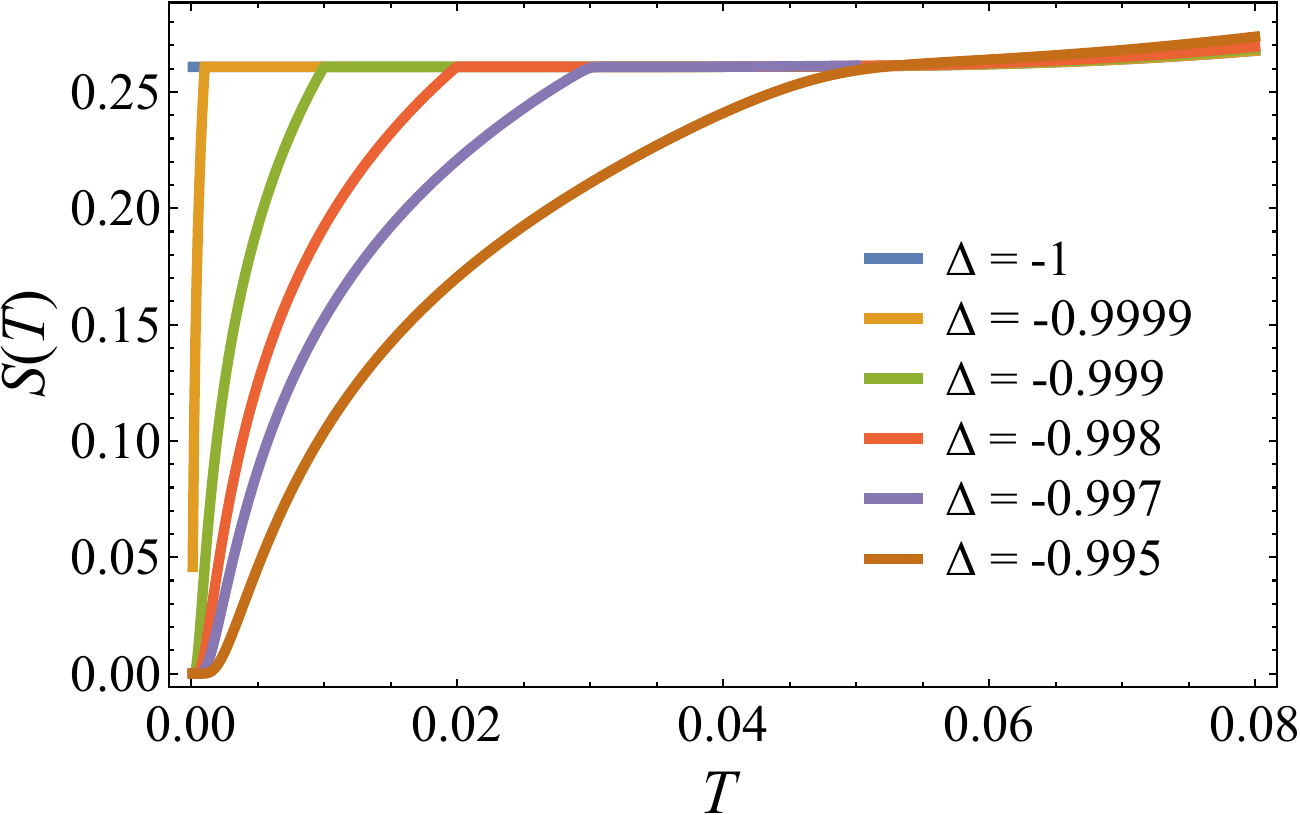}
	\end{minipage}
	\hfill
	\begin{minipage}{0.49\linewidth}
		\centering
		(b) Specific heat
		\includegraphics[width=\linewidth]{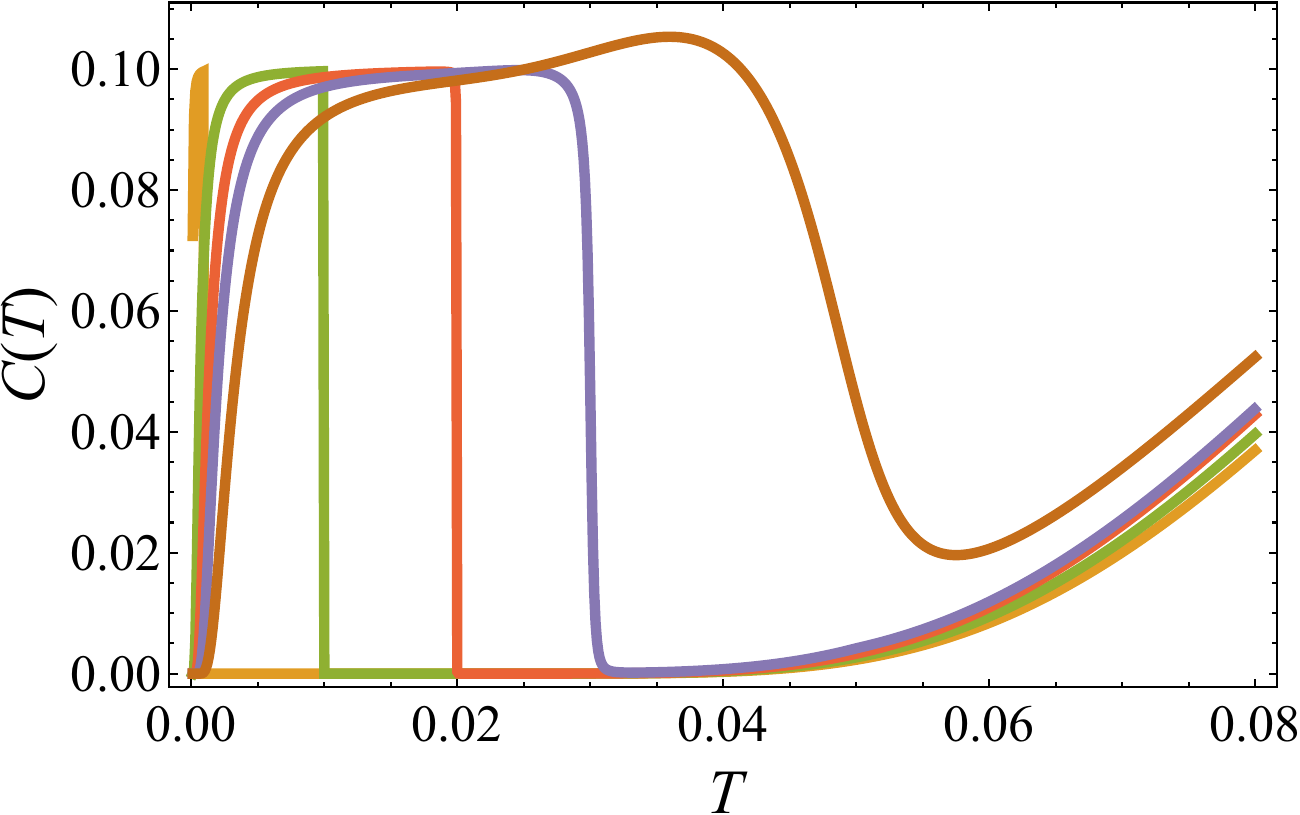}
	\end{minipage}
	\caption{Temperature dependences of (a) entropy and (b) specific heat demonstrate ``second-order'' pseudotransitions from the COI quasi-phase to the FM/AFM+PS quasi-phase with charge phase separation}
	\label{fig4}
\end{figure}

In this case, the pseudotransition occurs from the frustrated charge quasi-phase COI to the magnetic quasi-phase with phase separation FM/AFM+PS. However, the presence of phase separation leads to a continuous pseudo-transition, which notably resembles a second-order phase transition: the heat capacity changes abruptly, and the entropy has a sharp inflection (see Figure~\ref{fig4}). In this case, entropy gradually decreases as the system cools until it reaches zero. The charge and magnetic quasi-phases are partially compatible due to the presence of macroscopic phase separation. As a result, the high-entropy quasi-phase exists within the phase separation in the form of short-range order, which gradually disappears as the temperature decreases. This type of pseudotransition has not previously been discovered in frustrated systems.

This type of pseudotransition is fundamentally different from pseudotransitions found in other frustrated one-dimensional models. In this regard, we divide the two pseudotransitions discovered within this work into ''first-order`` pseudotransitions and ''second-order`` pseudotransitions.

\section{Acknowledgment}

This work was supported by the Ministry of Science and Higher Education of the Russian Federation, project FEUZ-2023-0017.
	
	\makeatletter


\addcontentsline{toc}{section}{References}
	\begin{thebibliography}{99}
		
		\bibitem{1} C. Coulon, H. Miyasaka, R. // Clérac Struct. Bond 122, 163 (2006).
		
		\bibitem{2} E. Aydıner et al. // Phys. Status Solidi B 243, 2901 (2006).
		
		\bibitem{3} S.M. de Souza, O. Rojas // Solid State Commun. 269, 131 (2017).
		
		\bibitem{4}	M.E. Zhitomirsky // Phys. Rev. B 67, 104421 (2003).
		
		\bibitem{5} O. Rojas // Acta Phys. Pol. A 137, 933 (2020).
		
		
	\end{thebibliography}
\end{document}